\documentclass{IAU-TrB}
\usepackage{graphicx, url, hyperref, natbib}
\newcommand{\NOPRINT}[1]{\null}
\def\apj{ApJ}%
\def\apjl{ApJ}%
\def\apjs{ApJS}%
\def\aap{A\&A}%
\def\mnras{MNRAS}%
\title[THEORY OF STELLAR ATMOSPHERES]     
{}

\author[DIVISION G COMMISSION 36]   
{}

\pubyear{2015}
\volume{Volume XXIXA}
\pagerange{1--4}
\date{15 Sep 2015}
\jname{Transactions IAU, Volume XXIXA}
\editors{Thierry Montmerle, ed.}
\begin{document}

\maketitle

{\bf

\large
\begin{tabbing}
\hspace*{65mm}       \=                                              \kill
DIVISION G \\COMMISSION 36         \> THEORY OF STELLAR ATMOSPHERES                                      \\
                     \> {\it (TH\a'EORIE DES ATMOSPH\a`ERES STELLAIRES)}                             \\
\end{tabbing}

\normalsize

\begin{tabbing}
\hspace*{65mm}       \=                                               \kill
PRESIDENT            \> Joachim Puls                            \\
VICE-PRESIDENT       \> Ivan Hubeny                                  \\
PAST PRESIDENT       \> Martin Asplund                           \\
ORGANIZING COMMITTEE \> France Allard, Carlos Allende Prieto,            \\
                     \> Thomas R. Ayres, Mats Carlsson,          \\
                     \> Bengt Gustafsson, Rolf-Peter Kudritzki     \\
                     \> Tatiana A. Ryabchikova        \\
\end{tabbing}

\bigskip
\noindent
HEXENNIAL REPORT 2009-2015
}

\small

\firstsection 
\section{Introduction}

\vspace{3ex} 
\noindent 
Different from previous triennial reports, this report covers the
activities of IAU Commission~36 `Theory of Stellar Atmospheres' over
the past {\it six} years\footnote{because of technical reasons, the report
on the years 2009 to 2012 could not be delivered.}, and will be the
last report from the `old' Commission~36. After the General Assembly
in Honolulu (August 2015), a new Commission `Stellar and Planetary
Atmospheres' (C.G5, under Division G, `Stars and Stellar Physics') has
come into life, and will continue our work devoted to the outer
envelopes of stars, as well as extend it to the atmospheres of planets
(see Sect.~4).  

From its establishment in 1970 on (with Commission President Richard Thomas),
Commission~36 has covered the field of the {\it physics} of stellar
atmospheres, and closely related topics. For all this time, and also
during the last six years, the scientific activities in this large
field have been very intense, and have led to the publication of a
large number of papers, which makes a detailed report even on the last
hexennium almost impossible. 

We have therefore decided on a two step approach. In the first part
(Sect.~5), we highlight specific contributions that directly refer to
the central topic of our -- now ending -- Commission, namely the {\it
theory} of stellar atmospheres. Because of the somewhat different
approaches, assumptions and methods, we divide this section into the
atmospheres of late-type (low and intermediate mass) stars, and of
massive stars. In the second part (Sect.~6), we keep the format of the
preceding reports, namely we list the areas of current research. Web
links for obtaining further information are provided in Sect.~7.

At first, however, we will briefly outline the composition of the
Organization Committee of Commission~36 during the triennium 2009 to
2012 in Sect.~2 (the members quoted at the beginning of this report
refer to the triennium between 2012 and 2015), summarize the scientific
meetings held between 2009 and 2015 that were relevant for our
Commission (Sect.~3), and comment on the establishment of the new
Commission C.G5 (comprising an extended scientific field compared to
the `old' one) within the re-structuring process of the IAU. 

\section{Presidents and OC of Commission~36 during the triennium
2009 to 2012}
\noindent
President: Martin Asplund; Vice-President: Joachim Puls; Past
President: John D. Landstreet; Organization Committee members:
Carlos Allende Prieto, Thomas R. Ayres, Svetlana V. Berdyugina, 
Bengt Gustafsson, Ivan Hubeny, Hans G. Ludwig, Lyudmila I. Mashonkina, 
Sofia Randich 

\section{Scientific meetings related to the interests of Commission~36}
\noindent
Many conferences and workshops have been held during the period
covered by this report on topics related to the interests of
Commission~36. The following symposia were sponsored by the IAU: 
IAU Symposium No. 265 {\it Chemical Abundances in the Universe: Connecting
First Stars to Planets};
IAU Symposium No. 268 {\it Light Elements in the Universe};
IAU Symposium No. 272 {\it Active OB stars: structure, evolution, mass
loss, and critical limits};
IAU Symposium No. 273 {\it The Physics of Sun and Star Spots};
IAU Symposium No. 279 {\it The Death of Massive Stars: Supernovae and
Gamma-Ray Bursts};
IAU Symposium No. 282 {\it From Interacting Binaries to Exoplanets:
Essential Modeling Tools};
IAU Symposium No. 283 {\it Planetary Nebulae: An Eye to the Future};
IAU Symposium No. 294 {\it Solar and Astrophysical Dynamos and Magnetic
Activity};
IAU Symposium No. 298 {\it Setting the scene for Gaia and LAMOST};
IAU Symposium No. 301 {\it Precision Asteroseismology};
IAU Symposium No. 302 {\it Magnetic Fields throughout Stellar Evolution};
IAU Symposium No. 305 {\it Polarimetry: The Sun to Stars and Stellar
Environments}; 	
IAU Symposium No. 307 {\it New windows on massive stars: asteroseismology,
interferometry, and spectropolarimetry}.

Our members also participated in many of the Joint Discussions and
Special Sessions at the IAU XXVII General Assembly in Rio de Janeiro,
August 2009 (e.g., 
JD4 {\it Progress in Understanding the Physics of Ap and Related Stars};
JD10 {\it 3D Views on Cool Stellar Atmospheres - Theory Meets Observation};
JD11 {\it New Advances in Helio- and Astero-Seismology};
JD13 {\it Eta Carinae in the Context of the Most Massive Stars}; 
SpS1 {\it IR and Sub-mm Spectroscopy - a New Tool for Studying Stellar
Evolution};
SpS7 {\it Young Stars, Brown Dwarfs, and Protoplanetary Disks}),\\
and at the IAU XXVIII General Assembly in Beijing, August 2012 (e.g., 
JD2 {\it Very massive stars in the local universe};
SpS5 {\it The IR view of massive stars: the main sequence and beyond};
SpS13 {\it High-precision tests of stellar physics from high-precision
photometry}).

Meetings not organised under the auspices of the IAU also attracted
interest. Among others, the following international meetings were
attended by our members:

{\it Deciphering the Universe through Spectroscopy}, September
2009, Potsdam, Germany;
{\it The 3rd Magnetism in Massive Stars (MiMeS) Workshop}, November 2009,
Hawaii, USA;
{\it Magnetic Fields: From Core Collapse to Young Stellar Objects}, May
2010, London, Ontario, Canada;
{\it Binary Star Evolution: Mass Loss, Accretion, and Mergers}, June 2010,
Mykonos, Greece;
{\it The Multi-Wavelength View of Hot, Massive Stars, July 2010}, Liege,
Belgium;
{\it The 10th International Colloquium on Atomic Spectra and Oscillator
Strengths for Astrophysical and Laboratory Plasmas}, August 2010,
Berkeley, USA; 
{\it 17th European White Dwarf Workshop}, August 2010, T\"ubingen, Germany;
{\it The 16th Cambridge Workshop on Cool Stars, Stellar Systems and the
Sun}, September 2010, Seattle, USA;
{\it Workshop on Convection in Stars}, January 2011, Johannesburg,
South-Africa;
{\it 8th Serbian Conference on Spectral Line Shapes in Astrophysics}, June
2011, Divcibare, Serbia;
{\it Stellar Atmospheres in the Gaia Era: Quantitative Spectroscopy and
Comparative Spectrum Modelling}, June 2011, Brussels, Belgium;
{\it Four Decades of Research on Massive Stars: A Scientific Meeting in the
Honour of Anthony F. J. Moffat}, July 2011, Saint-Michel-des-Saints,
Canada;
{\it The Fifth Meeting on Hot Subdwarf Stars and Related Objects}, July 2011,
Stellenbosch, South Africa;
{\it From Atoms to Stars: The Impact of Spectroscopy on Astrophysics}, July
2011, Oxford, UK;
{\it The Mass Loss Return from Stars to Galaxies}, March 2012, Baltimore, USA;
{\it Circumstellar Dynamics at High Resolution}, March 2012, Foz do Iguacu, Brazil 
{\it 30 Doradus: The Starburst Next Door, September 2012}, Baltimore, USA;
{\it 50 Years of Brown Dwarfs: from Theoretical Prediction to Astrophysical
Studies}, October 2012, Ringberg Castle, Germany;
{\it Putting A Stars into Context: Evolution, Environment, and Related
Stars}, June 2013, Moscow, Russia;
{\it Massive Stars: From alpha to Omega}, June 2013, Rhodes, Greece;
{\it 11th International Colloquium on Atomic Spectra and Oscillator
Strengths for Astrophysical and Laboratory Plasmas}, August 2013, Mons,
Belgium;
{\it 400 Years of Stellar Rotation}, November 2013, Natal, Brazil;
{\it Ionising processes in atmospheric environments of planets, Brown
Dwarfs, and M-dwarfs}, January 2014, RAS, London, UK;
{\it The 18th Cambridge Workshop on Cool Stars, Stellar Systems and the
Sun}, June 2014, Flagstaff, USA;
{\it Fast outflows in massive stars: from single objects to wind-fed and
colliding-wind binaries}, July 2014, Geneva, Switzerland;
{\it X-ray Astrophysics of Hot Massive Stars}, Scientific Event E1.3,
COSPAR-14, August 2014, Moscow, Russia;
{\it Bright Emissaries: Be Stars as Messengers of Star-Disk Physics}, August
2014, London, Ontario, Canada;
{\it Magnetism and Variability in O stars}, September 2014, Amsterdam, The
Netherlands;
{\it The Brown Dwarf to Exoplanet Connection Conference: Making sense of
Atmospheres and Formation}, October 2014, Newark, Delaware, USA;
{\it Ringberg Workshop on Spectroscopy with the Stratospheric Observatory
For Infrared Astronomy}, March 2015, Ringberg Castle, Germany;
{\it Massive Stars and the Gaia-ESO Survey}, May 2015, Brussels, Belgium;
{\it International Workshop on Wolf-Rayet Stars}, June 2015, Potsdam, Germany;
{\it The physics of evolved stars: A conference dedicated to the memory of
Olivier Chesneau}, June 2015, Nice, France.

\section{Towards a new Commission on `Stellar and Planetary Atmospheres'}
\noindent
During the General Assembly (GA) in Beijing 2012, several steps had
been decided upon to initiate a re-structuring process of the IAU
scientific bodies. At first, new Divisions were created that came
into live just after the GA (regarding Commission~36, this refers to
our new parent-Division G `Stars and Stellar Physics'). Within a
second step, all old Commissions were planned to terminate with the
upcoming GA in Honolulu 2015, with new Commissions to be
suggested and applied for during the year 2014. To this end, a
subgroup of our OC, together with other interested scientists, were
installed to discuss and prepare a proposal for a new Commission
related to our present subjectives. After many discussions, a final
group chaired by Ivan Hubeny and co-proposers France Allard, Katia
Cunha and Adam Showman submitted a proposal on the establishment of a
new Commission on `Stellar and Planetary Atmospheres' within Division
G. The original proposal (including the ideas/objectives behind it)
can be found at {\tt
http://www.iau.org/submissions/newcommissions/detail/81/}, where it
was suggested to 

``{\ldots} reshape the original Commission 36, `Theory of Stellar
Atmospheres', and to extend its scope beyond the stars to atmospheres
of substellar objects like Brown Dwarfs and extrasolar planets.''

In brief, it was thought that such an extension is possible and {\it
required} in the present astronomical context, since the latter
objects and their atmospheres are in the focus of intense studies (now
and in future), and the physics/techniques to model and study their
atmospheres are quite similar to the approach performed in the
investigations of stellar atmospheres. The proposers concluded that 
``a fruitful interaction between the stellar and planetary
communities, with connections to other areas where the stellar
modeling paradigm can be applied, will be mutually beneficial.''

Indeed, this proposal was approved by the IAU Executive Committee
(Padua, Italy, 15-17 April 2015), together with various other
proposals. Subsequently, the IAU members were asked to decide whether
and in which Commission they want to participate, and 207 (status July
2015) decided to become a member of the new Commission C.G5. The new
Commission President is the chair of the proposers, and the
co-proposers became members of the OC. Just before the GA in Honolulu, 
elections for the Vice-President and the remaining OC members were
conducted, and the
complete list for our new Commission `Stellar and Planetary
Atmospheres' reads as follows:

President: Ivan Hubeny, Vice-President: Carlos Allende Prieto,
OC-members: France Allard, Katia Cunha, John D. Landstreet, Thierry
Lanz, Lyudmila I. Mashonkina, Adam Showman.

We are convinced that the new Commission (which started its activities
just after the GA in Honolulu) will successfully continue and extend
our previous work, and that they will promote the physics and
modeling of stellar and planetary atmospheres worldwide. 

\section{Specific contributions related to the {\it theory} of stellar
atmospheres}
\noindent
Before we summarize specific contributions and achievements within our
research field during the last hexennium -- as outlined in the
introduction, here we will concentrate on theoretical work
-- let us mention that during the period covered by this report, a
major textbook on stellar atmospheres was written and published, by
Ivan Hubeny and the late\footnote{Dimitri Mihalas, known to all of us
as a world-leading expert on stellar atmospheres, radiation hydrodynamics
and spectroscopy, passed away on Nov. 21, 2013. We will greatly miss him.}
Dimitri Mihalas \citep{HM14}, who also served as a President of this
Commission between 1976 to 1979. In this textbook, one can find over
1200 references, many referring to publications from the covered time
period.

\subsection{Late-type stars}

\paragraph{\it 1D hydrostatic atmosphere models.}
One-dimensional, hydrostatic atmosphere models of late-type stars continue
to be a staple food in the field of stellar physics and beyond in astronomy.
Such models are continuously improved, especially in terms of accuracy and completeness
of input data such as opacities. In recent years, extended grids of 1D, LTE stellar
atmosphere models across a wide
range of stellar parameter space for AFGKM stars have been computed using
the {\sc marcs} 
\citep{2008A&A...486..951G, 
2012AJ....144..120M}, 
{\sc mafags} 
 \citep{2009A&A...503..177G} 
 and {\sc phoenix} 
 \citep{2013A&A...553A...6H} 
 codes. 
 \citet{2009ApJ...691.1634S} 
 investigated the importance of non-LTE effects for the atmospheric structures and
 UV radiation fields for a few benchmark stars, concluding that shortcomings in atomic data still exist.
    \citet{2012A&A...546A..14C, 2013A&A...552A..16C} 
 have employed 1D {\sc phoenix} models to calculate detailed limb-darkening coefficients
 for a variety of photometric systems.
 \citet{2012A&A...544A.126D} 
calculated detailed synthetic stellar spectra for the {\sc marcs} grid of models and for
a variety of chemical compositions to be used for stellar parameter and abundance determinations,
especially for the new generation of large-scale spectroscopic surveys, such as the 
Gaia-ESO 
 \citep{2012Msngr.147...25G} 
and
the GALAH survey 
 \citep{2015MNRAS.449.2604D}. 

\smallskip
\noindent
\paragraph{\it  Stellar winds and mass loss of red giants and AGB stars.}
Models of dust-driven winds of carbon stars have been computed by 
 \citet{2010A&A...509A..14M}, 
 \citet{2011A&A...533A..42M}, 
 and 
  \citet{2014A&A...566A..95E}, 
including an investigation of the importance of grain size on the resulting wind properties. 
Similar modelling but for M-type asymptotic giant branch stars has been carried out by
 \citet{2012A&A...546A..76B}, 
 who found that a two-stage process is required: 
 atmospheric levitation by pulsation-induced shock waves followed by radiative acceleration on dust grains. 
 They investigated possible dust species, ruling out most but speculated that Mg$_2$SiO$_4$ is the main actor,
 which  was confirmed by more detailed calculations by 
 \citet{2015A&A...575A.105B}. 

\smallskip
\noindent
\paragraph{\it  3D hydrodynamical models of stellar surface convection and atmospheres.}
More physically motivated and realistic models of the atmospheres and surface convection
of late-type stars are now possible to compute using full 3D, time-dependent, radiative-hydrodynamical
simulations that self-consistently predict the crucial radiative heating/cooling and convective
energy transport 
 \citep[e.g.][and references therein]{2009LRSP....6....2N}. 
Recently such modelling has been extended to stars other than the Sun, including
for solar-type stars
 \citep{2013A&A...558A..48B, 
 2010A&A...517A..49H}, 
red giant stars
 \citep{2015A&A...576A.128D}, 
 AGB and supergiants
 \citep{2011A&A...535A..22C}, 
metal-poor stars
 \citep{2011A&A...528A..32C}, 
white dwarfs
 \citep{2013A&A...557A...7T}, 
 M dwarfs 
  \citep{2009A&A...508.1429W}, 
 brown dwarfs
  \citep{2010A&A...513A..19F}, 
and Cepheids
 \citep{2015MNRAS.449.2539M}. 
Several 3D hydrodynamics codes are available for the purpose, which have
been used to compute extensive grids of realistic 3D stellar models, including
{\sc stagger}
 \citep{2013A&A...557A..26M, 
 2013A&A...560A...8M}, 
{\sc co5bold}
\citep{2012JCoPh.231..919F, 
2012A&A...547A.118L}, 
{\sc muram}
\citep{2005A&A...429..335V, 
2009ApJ...691..640R}, 
{\sc antares}
 \citep{2010NewA...15..460M, 
2015NewA...34..278G}, 
and the 
 \citet{1998ApJ...499..914S} 
 code 
 \citep{2013ApJ...769...18T}. 
 \citet{2012A&A...539A.121B} 
concluded that different codes produce very similar results for the case of the Sun.
  \citet{2013A&A...554A.118P} 
carried out a detailed comparison of the predictions of a 3D solar model computed
with the {\sc stagger} code against an arsenal of observational diagnostics and found
extremely satisfactory agreement, demonstrating that such 3D models are indeed highly realistic.
Such 3D stellar models have been used among others to calculate 
limb-darkening
 \citep{2015A&A...573A..90M}, 
 stellar spectra
  \citep{2009A&A...501.1087R}, 
and stellar oscillations and asteroseismic diagnostics
\citep{2013A&A...559A..39S, 
2013A&A...559A..40S}. 

\smallskip
\noindent
\paragraph{\it  Stellar magneto-convection.}
Magnetic fields are ubiquitous in late-type stars such as the Sun
 \citep[e.g.][and references therein]{2012LRSP....9....4S}. 
  \citet{2010ApJ...724.1536F} 
  and 
   \citet{2012A&A...548A..35F} 
calculated 3D MHD models of varying magnetic field strengths for the Sun
and investigated the impact on the resulting spectral line profiles and inferred
solar Fe abundance, finding significant differences compared with pure
3D hydrodynamical models. However, 
 \citet{2015ApJ...799..150M} 
found that the assumed magnetic field topology is important and concluded that
more realistic initial configurations result in reduced differences with previous, pure hydrodynamical
modelling.
 \citet{2015A&A...581A..42B} 
 have calculated 3D MHD models  for different FGM dwarfs for a range of field strengths, discovering notable
 differences in the manifestation of the magnetic fields depending on the stellar parameters. 
 \citet{2015A&A...581A..43B} 
 used these 3D MHD models to investigate the impact on the predicted stellar spectra. 
The physics and dynamics of sunspots have been modelled by 
\citet{2009Sci...325..171R} 
and
\citet{2011ApJ...740...15R} 
by means of 3D MHD simulations, achieving impressive agreement with observations. 
\citet{2015ApJ...803...42H} 
and 
\citet{2014ApJ...789..132R} 
have investigated the small-scale dynamo in the solar convection zone and how it generates the
solar magnetic field and its various manifestations.

\smallskip
\noindent
\paragraph{\it  Calibrating the mixing length theory using 3D stellar models.}
One very attractive feature with realistic 3D hydrodynamic stellar models is
their ability to predict the convective energy transport without invoking any free parameters
such as the traditional mixing length parameters or close relative thereof required in 1D 
stellar atmosphere and interior models. 
It is thus possible to calibrate the mixing length theory using 3D models, something 
which has recently been carried out by 
 \citet{2014MNRAS.445.4366T} 
 and
 \citet{2015A&A...573A..89M}. 
They find significant variations of the mixing length parameter across the HR-diagram,
in contrast to the constant value calibrated on the Sun invariably assumed in traditional 1D modelling.

\smallskip
\noindent
\paragraph{\it Non-LTE radiative transfer.}
Detailed radiative transfer calculations taking into account departures from local thermodynamic equilibrium
for key elements and transitions have been carried out by several groups, including by  
\citet{2013ApJ...764..115B} 
who investigated line formation in red supergiants. 
\citet{2011A&A...528A..87M} 
and 
\citet{2012MNRAS.427...50L} 
performed comprehensive non-LTE computations for Fe for late-type stars using 1D model atmospheres. 
\citet{2012MNRAS.427...27B} 
did similar calculations for a selection of well-studied benchmark stars but employed also
spatially and temporally averaged 3D models (thus reducing them to effectively
1D models but with a more realistic atmospheric stratification). 
Full 3D non-LTE calculations  have been carried out by 
\citet{2015MNRAS.454L..11A}, 
who investigated the line formation of O across a large range of stellar parameter space, finding
substantial non-LTE effects at solar metallicity but surprisingly small departures at low metallicity.
\citet{2015arXiv150803487S} 
have performed 3D non-LTE line formation calculations for the O\,{\sc i} 777nm triplet in the Sun.
\citet{2013A&A...554A..96L}, 
and 
\citet{2010A&A...522A..26S} 
have investigated departures from LTE in 3D hydrodynamical stellar atmosphere models for Li in
metal-poor stars. 
\citet{HausBaron10}, 
\citet{HausBaron14} 
and
 \citet{2015A&A...574A...3P} 
have developed new and computationally efficient codes to handle 3D non-LTE radiative transfer
for parallel processing (see also Sect.~\ref{massivestars}).

\smallskip
\noindent
\paragraph{\it Solar chemical composition.}
The exact chemical make-up of the Sun continues to attract a great deal of attention.
This fundamental yardstick for astronomy has undergone a drastic downward revision
for the most abundant metals -- C, N, O and Ne in particular -- over the past 15 years.
 \citet{2009ARA&A..47..481A}
 published a comprehensive solar analysis of all spectroscopically accessible elements
 using a realistic 3D hydrodynamic solar atmosphere {\sc stagger} model, non-LTE spectral line formation,
 updated atomic/molecular data and careful consideration of blending lines. 
 This is the first time all elements have been analysed homogeneously and with 
 the statistical and systematic uncertainties carefully estimated. 
Further details of their 3D-based solar analysis are available in
 \citet{2009A&A...507..417P}, 
  \citet{2015A&A...573A..25S}, 
 \citet{2015A&A...573A..26S}, 
 and
  \citet{2015A&A...573A..27G}. 
A similar solar analysis using an independent 3D solar model computed with the
{\sc co5bold} code but for a restricted selection of elements has been carried out 
by  
\citet{2009A&A...498..877C}, 
\citet{2010A&A...514A..92C} 
and
 \citet{2015A&A...579A..88C}; 
a summary of the group's findings in terms of the solar composition is available in
\citet{2011SoPh..268..255C}. 
Their derived C, N and O abundances are intermediate between the low values advocated by 
 \citet{2009ARA&A..47..481A}
and the canonical high values from two decades ago. 
Similar intermediate abundances are supported by 
 \citet{2009ApJ...704.1174P} 
who assessed the available spectroscopic analysis at the time although did not carry out their own 
line formation calculations. 
  \citet{2010ApJ...724.1536F} 
and 
 \citet{2015ApJ...799..150M} 
have investigated the influence of magnetic fields in the quiet Sun on the emergent spectral line profiles 
and derived abundances of Fe, concluding that the impact is very modest for the transitions
typically used in solar abundance analysis. 

The new solar chemical composition, especially the low C, N, O and Ne abundances found by
 \citet{2009ARA&A..47..481A}, 
has caused a great deal of consternation 
within the helioseismology 
community: solar models constructed with the new solar chemical composition 
no longer agree with the sound speed variation with depth, He abundance
in the convection zone and the depth of the 
convection zone inferred from solar oscillations 
 \citep[e.g.][]{2009ApJ...705L.123S, 
2014ApJ...787...13V}. 
Many possible solutions to this {\em solar modelling problem} have been proposed over the past decade, 
none entirely successful to date. The suggestion that the problem arises from missing opacity in the solar
interior, especially immediately below the convection zone, has recently received very convincing support
from new experimental opacity measurements 
 \citep{2015Natur.517...56B}. 
These new data imply that the predicted opacities for Fe based on
state-of-the-art calculations such as OP and OPAL are wrong by
30-400\%. By themselves, the new Fe opacities can explain half of the
{\em solar modelling problem}; similar experiments for other elements
and improved atomic physics calculations are urgently needed whether
missing opacity is the full solution. It is important to note that
this is a {\em stellar} physics problem rather than only facing {\rm
solar} physics: whatever the solution turns out to be, it will have
profound implications for all of stellar physics and by inference much
of astronomy. 

\smallskip
\noindent
\paragraph{\it First stars.}
Much work in recent years has focussed on the nature of the very first generations of stars
formed after the Big Bang as inferred from the chemical compositions of extremely metal-poor stars. 
Detailed 3D hydrodynamical stellar atmosphere models have been calculated for a few of the
most metal-poor stars discovered to date, including for SMSS J031300.36-670839.3, the current 
record-breaker in terms of Fe abundance
 \citep{2014Natur.506..463K}: 
a remarkable upper limit of $10^{-7.1}$ times the solar abundance. 
This star however has enormous over-abundances of C and O as inferred from
3D LTE calculations of CH and OH lines
 \citep{2015ApJ...806L..16B}, 
making its overall metal content low but not particularly extreme. 
Instead the star with the lowest overall metallicity is SDSS
J102915+172927, for which 
 \citet{2011Natur.477...67C} 
analysed the chemical composition using a 3D stellar atmosphere
{\sc co5bold} model. 

\smallskip
\noindent
\paragraph{\it Molecule and dust formation in brown dwarfs and (exo-)planetary atmospheres.}
The low temperatures and high densities encountered in the atmospheres of brown dwarfs and exoplanets 
are conducive to the formation of complex molecules and eventually nucleation and condensation of various
species of dust
 \citep[e.g.][]{2012RSPTA.370.2765A, 
2013RSPTA.37110581H}. 
 \citet{2014A&ARv..22...80H} 
has reviewed the literature of atmospheres of brown dwarfs, remarking
that dust forms already at effective temperatures of about 2,800\,K, coincidentally marking roughly the boundary 
between M dwarfs and brown dwarfs. 
 \citet{2009A&A...506.1367W} 
performed detailed calculations for dust nucleation, condensation and evaporation and allowing for dust drift in brown dwarfs and extra-solar planets
for a range of stellar parameters and chemical compositions using 1D {\sc phoenix} models. They concluded that dust formation
is ubiquitous and that
the dust-to-gas ratio does not scale linearly with the object's metallicity for a given effective temperature.
 \citet{2010A&A...513A..19F} 
 have studied the role of convection, overshoot, and gravity waves for the transport of dust in M dwarf and brown dwarf atmospheres
 using 3D hydrodynamical models with a simplified treatment of dust formation.

\subsection{Massive stars and related objects}
\label{massivestars}

\paragraph{\it Atmospheric models/spectrum synthesis} 
for massive stars require(s) a non-LTE (NLTE) treatment, and in many
cases the inclusion of a stellar wind. \citet{Przybilla10a} summarized
the construction of the underlying model atoms, and \citet{Puls09}
reviewed various aspects of the above requirements, including a brief
comparison of the most-used model codes suited for the analysis of
massive stars, namely DETAIL/SURFACE, TLUSTY/SYNSPEC (both codes:
plane-parallel, hydrostatic), CMFGEN, FASTWIND, PHOENIX, PoWR, and
WM-{\sc basic} (latter five codes: spherical, including winds). For
original references, see \citet{Puls09}, but note that most of these
codes have been updated since release, which is also true for the NLTE
OB-star wind models updated by \citet{KK09} regarding the X-ray
emission from wind-embedded shocks. Overviews of specific codes can be
found in \citet[TLUSTY]{HubenyLanz11a},
\citet[SYNSPEC]{HubenyLanz11b}, \citet[CMFGEN]{Hillier12},
\citet[FASTWIND]{Rivero12} and \citet[PoWR]{Hamann09}. Model grids for
massive stars have published by \citet[CMFGEN]{Zsargo13} and by
\citet[PoRW: WN-stars]{Todt15}. \citet{Massey13} compared the results
when analyzing observed spectra by means of either CMFGEN or FASTWIND,
whilst \citet{Przybilla11} compared the results from ATLAS9/SYNTHE and
TLUSTY/SYNSPEC vs. DETAIL/SURFACE.

Similar (NLTE-) methods as used above have been applied to model
the ejecta/remnants of supernovae and corresponding
emergent spectra (\citealt{Noebauer12, KerzendorfSim14, Pauldrach14}),
to calculate theoretical models/spectra of accretion disks in AGN
\citep{Hubeny10}, and to model the atmospheres of Brown Dwarfs and extrasolar
giant planets \citep{Hubeny12}. 

\smallskip
\noindent
\paragraph{\it Multi-D radiative transfer.}
The `general-purpose' code PHOENIX has been updated and tested for 3-D
radiative transfer in spherical/cylindrical coordinate systems and an
operator-splitting technique \citep{HausBaron09, HausBaron10},
including a multi-level NLTE description \citep{HausBaron14}.
\citet{Weber13} reported on 3-D modeling of ionized gas (objective:
cosmic reionization) and the 3-D, time-dependent modeling of the
metal ionization in HII regions, irradiated by consistent atmospheric
models of (very) massive stars. \citet{Ibgui13} presented a first
version of the 3-D, plane-parallel radiative transfer code IRIS.

\smallskip
\noindent
\paragraph{\it Time-dependent radiative transfer}
has been implemented in the context of spectrum synthesis of
supernovae \citep{KromerSim09, HillierDes12, Dessart14, DessartHil15,
Dessart15}.

\smallskip 
\noindent 
\paragraph{\it Macroturbulence and subsurface convection.} 
As a first hypothesis, supersonic macroturbulence (detected in the
spectral lines from the majority of OB-stars) has been suggested as a
collective effect from gravity mode pulsations \citep{Aerts09}, and
such relation has been studied by, e.g., \citet{Simon10, Simon11b,
Simon12}. Macro- (and also micro-)turbulence might be also related to
the presence of a subsurface convection zone due to the iron-opacity
peak \citep{Cantiello09, Grassitelli15}. Such convection zones might
be suppressed by strong magnetic fields, as indicated by absent
macroturbulence in the highly magnetic star NGC 1624-2
\citep{Sundqvist13}. 

\smallskip 
\noindent 
\paragraph{\it Envelope inflation.} 
Also because of the iron-opacity peak, stellar envelopes might become
inflated near the Eddington-limit, which would explain the (previous)
discrepancy between the radii of Wolf-Rayet stars derived either from
spectroscopy or from stellar modeling \citep{Graefener12}. Such
inflation has meanwhile been studied also in evolutionary models, both
for massive main-sequence stars \citep{Sanyal15a} and for Wolf-Rayet
stars \citep{Sanyal15b}. 

\smallskip 
\noindent 
\paragraph{\it Stellar winds and outflows -- stationary mass-loss.} 
A new method to model line-driven winds, based on Monte Carlo
radiation hydrodynamics, has been presented by \citet{Noebauer15}.
\citet{Cure11} studied `slow' wind solutions for A-type Supergiants,
and \citet{Silaj14} revisited line-driven winds in the context of Be
Stars, based on such slow solutions. \citet{MuellerVink14} provided 
solutions for the velocity field and mass-loss rates for 2-D
axisymmetric outflows, whilst in a series of papers \citet{Lucy10a,
Lucy10b, Lucy12a} derived O-star mass fluxes (also at low
metallicities) using a code for constructing moving reversing layers. 

In the context of so-called weak winds, \citet{Lucy12b} revised a
phenomenological two-component (hot and cool gas) model, where the
external outflow turns out to become a decelerating, coronal wind (see also
\citealt{Huene12} below).

In rapidly rotating stars close to critical rotation, mass and angular
momentum can be lost via decretion disks \citep{Krticka11}, and
time-dependent models of such disks have been modeled by
\citet{Kurfuerst14}.

\smallskip 
\noindent 
\paragraph{\it Stellar winds -- mass-loss near the Eddington limit and
Very Massive Stars (VMS),}
Wind models of VMS have been presented by \citet{Vink11} and
\citet{Pauldrach12}, where the former authors concentrated on optically thick
and the latter on optically thin winds. \citet{Graefener11} stressed
that the Eddington factor is key to understand the winds of the most
massive stars, and found evidence for an Eddington-Gamma dependence of
Wolf-Rayet type mass loss. The sub-photospheric layers of classical
Wolf-Rayet stars were traced by \citet{GrafVink13}, and
\citet{OwoShaviv12} provided their view on instabilities and mass
loss near the Eddington Limit.

\smallskip 
\noindent 
\paragraph{\it Stellar winds -- bi-stability braking.}
A potentially decisive process for the late and post main
sequence evolution of massive stars has been identified by
\citet{Vink10}, the so-called bi-stability braking: due to the
predicted increase of mass-loss over the bi-stability jump,
significant angular momentum might be lost, and the steep drop in the
rotation rates of B-supergiants below 22,000~K might be elegantly
explained.

\smallskip 
\noindent 
\paragraph{\it Stellar winds -- interaction with magnetic fields.}
Though only present in roughly 10\% of massive stars, magnetic fields
need to explained\footnote{to date, the hypothesis of fossil origin
is favoured, e.g., \citealt{Braithwaite14}}, and their interaction
with line-driven winds to be (further) studied. \citet{udDoula09}
performed dynamical simulations of magnetically channelled,
line-driven winds, and calculated the angular momentum loss and
rotational spin-down. \citet{Sundqvist12b} devised a dynamical
magnetosphere model for the periodic H$_\alpha$-emission from the
magnetic O star HD\,191612, and \citet{udDoulaetal13} performed first
3-D MHD simulations of a massive-star magnetosphere, applying their
model to the H$_\alpha$-emission from $\Theta^1$~Ori~C. 
The radiative cooling in multi-D models of magnetically confined wind
shocks was investigated by \citet{udDoula13}, and a magnetic
confinement versus rotation classification of massive-star
magnetospheres has been presented by \citet{Petit13}.

\smallskip 
\noindent 
\paragraph{\it Stellar winds -- instabilities/wind-embedded shocks.}
The possibility to obtain clumping also in the inner winds of hot,
massive stars (due to the presence of limb-darkening when calculating
the line-force) has been studied and discussed by \citet{SundOwo13}.
Such early onset has been actually measured, by means of X-ray
spectroscopy, in the wind of QV~Nor (B0I) \citep{Torre15}.
Progress on the effects of scattering (w.r.t. the transonic
solution topology and the intrinsic variability of line-driven winds)
has been obtained by \citet{SundOwo15}. \citet{Owo13} studied the
effects of thin-shell mixing in radiative wind-shocks, and
\citet{Cohen14b} measured the shock-heating rates in O-star winds
using X-ray line spectra.

\smallskip 
\noindent 
\paragraph{\it Stellar winds -- X-ray and Gamma-ray emission
(general, and from clumped winds in High Mass X-ray Binaries).}
The X-ray emission from wind-embedded shocks was studied by means of 
hydrodynamical simulations in NLTE wind models \citep{Krticka09}, and
\citet{Leuten10} modeled the broadband X-ray absorption in massive
star winds. An interesting possibility to solve the so-called weak
wind problem in massive stars by means of X-ray spectroscopy has been
presented by \citet{Huene12}, revealing a massive {\it hot} wind (for
the example of $\mu$~Col).

\citet{Naze14} investigated general aspects of the X-ray emission from
{\it magnetic} massive stars, whilst \citet{udDoulaetal14} considered
the specific effects of a cooling-regulated shock retreat w.r.t. the
X-ray emission from magnetically confined wind shocks. 

\citet{Oskinova12} highlighted the impact of wind clumping in
supergiant High Mass X-ray Binaries (HMXBs) on X-ray variability and
photoionization. Wind clumping also affects the variability at even
higher, Gamma-ray energies, in HMXBs with
jets \citep{Owo09}, and such Gamma-Ray Emission has been modeled by
\citet{Owo12} for the case of the HMXB LS~5039.

\smallskip 
\noindent 
\paragraph{\it Radiative transfer -- wind-inhomogeneities.} 
During the covered period, a main focus of radiative transfer in
massive star atmospheres was the (phenomenological) description and
implementation of wind-inhomogeneities (most likely related to the
line-driven instability). Particularly, the effects from optically
thick clumps leading to `macro-clumping' and porosity in physical and
velocity space have been investigated. The multi-D resonance line
formation in such inhomogeneous media was studied by
\citet{Sundqvist10} and \citet{Surlan12}, and macro-clumping/porosity was
suggested as a solution for the long-standing discrepancy between
H$_\alpha$ and P\,{\sc v} mass loss diagnostics \citep{Oskinova07,
Sundqvist11, Surlan13}. \citet{Sundqvist14} suggested an effective-opacity
formalism for the line transfer in accelerating, clumped two-component
media, which is ready to be implemented in state-of-the-art, NLTE
model atmosphere codes.

\smallskip 
\noindent 
\paragraph{\it Radiative transfer -- X-ray line diagnostics.} 
Another focus of radiative transfer was set to X-ray line
diagnostics, also here in combination with wind-inhomogeneities.
\citet{Sundqvist12a} formulated a generalized porosity formalism for
isotropic and anisotropic effective opacity (spherical vs.
oblate/prolate clumps), and studied its effects on the X-ray line
attenuation in clumped winds. \citet{Leuten13} provided constraints on
porosity and mass loss, from modeling of X-ray emission line profile
shapes, concluding that `X-ray mass-loss rate estimates are relatively
insensitive to both optically thin and optically thick clumping'.
Additionally, they compared their method with the alternative one
by \citet{Oskinova06}. 

Using various techniques (partly related to those just mentioned), the
X-ray spectra of individual Galactic O-stars were analyzed, and
mass-loss rates, structure and shock physics constrained
(\citealt{Cohen10} and \citealt{Herve13}: $\zeta~Pup$; \citealt{Rauw15}:
$\lambda~Cep$; \citealt{Cohen14a}: sample of Galactic O-stars).

\smallskip 
\noindent 
\paragraph{\it Radiative transfer -- The formation of specific lines}
affected by complex NLTE effects was studied in various publications. 
\citet{Rivero11} revisited the Nitrogen III emission line formation in
O-stars (those classified by `f'), and demonstrated that the emission
is due to wind effects, but rather unaffected by dielectronic
recombination as had been suggested previously. The formation of C\,{\sc iii}
4647-50-51 and C\,{\sc iii} 5696 in O star atmospheres, essential to
derive carbon abundances, has been explained by \citet{MartinsHil12}.
In this context, strong constraints on the [N/C] vs. [N/O] abundance
ratios (for CNO cycled material mixed into the atmosphere) that are
(almost) independent on specific evolutionary scenarios have been outlined
by \citet{Przybillaetal10}. 

\citet{Petrov14} discussed the formation and behaviour of H$_\alpha$
over the theoretically predicted bi-stability jump in blue supergiants,
and \citet{Przybilla10b} and \citet{Najarro11} investigated the NLTE
line formation in the near-IR and the L-band, respectively.

\smallskip 
\noindent 
\paragraph{\it Analysis techniques.}
Various publications dealt with specific analysis techniques. First,
we refer to problems in the application of the popular Fourier
transform method to derive rotational velocities, when analyzing
slowly rotating stars \citep{Sund13} or stars with time-dependent
profiles (because of pulsations, \citealt{Aerts14}). \citet{Simon11a}
presented a grid-based automatic tool for the quantitative
spectroscopic analysis of O-stars (IACOB-GBAT), whilst DISKSPEC
\citep{Hubeny13} is a new tool for analyzing the spectra of accretion
disk systems. BONNSAI \citep{Schneider14} is a Bayesian tool for
comparing observationally derived stellar parameters with stellar
evolution models. In this context, a valuable, radius-free analogue
to the conventional Hertzsprung-Russell diagram is provided by the
the `spectroscopic Hertzsprung-Russell diagram', with axes T$_{\rm
eff}^4/g$ vs. T$_{\rm eff}$ \citep{LangerKud14}.

\section{Primary research areas 2009 - 2015}
\noindent
\subsection{Physical processes}
\noindent
{\it General properties.} Line-blanketed, unified NLTE models of
massive star atmospheres (including winds) available to the community.
3D radiative-hydrodynamical models of stellar surface convection and atmospheres
for a range of temperatures, gravities and metallicities corresponding to late-type stars. 
Magneto-convection in stars. 
1D hydrodynamical models of atmospheres and dust-driven
winds of AGB stars. 
Dust nucleation and condensation in planetary and brown dwarf atmospheres. 
Continuous and atomic/molecular line opacities. 
Non-equilibrium spectrum formation and chemistry. 
Grids of synthetic fluxes and spectra. Calibrating parameterized
models through physical modeling. Calibrating abundance determinations
by filter photometry or low-resolution spectroscopy. Laboratory
studies of laser-induced plasma to simulate physical conditions in
stellar atmospheres.

{\it Hydrodynamical processes within stellar atmospheres.} Convection
(granulation) in surface layers, and its effects upon emergent
spectra. Interplay between convection and non-radial pulsation. Scales
of surface convection in stars in different stages of evolution.
Hydrodynamic simulations of entire stellar volumes. 
3D convection simulations of white dwarfs. 
Calibration of mixing length theory of convection using 3D hydrodynamical stellar models. 
Effects of magnetic fields on atmospheric structure and emergent spectrum. 
Excitation of solar/stellar oscillations by surface convection. Sub-surface
convection in massive stars. Inflation of massive star atmospheres.

{\it Transient processes.} Shocks in pulsating stars. Radiative
cooling of shocked gas. Emission lines as shock-wave diagnostics.
Co-rotating interaction regions in radiation driven stellar winds.
Particle acceleration during flares. Interaction of jets with
interstellar medium. Episodic outflows and star-disk interaction.

{\it Magnetic phenomena.} Magnetic structures in single and binary
stars. Dynamo generation of magnetic fields by convection. 
Observational manifestation of stellar magnetic fields. 
Magnetic surface phenomena: sun-/starspots, bright pores, 
active regions, flares etc. 
Magnetic and acoustic heating of solar/stellar chromospheres and coronae. 
Generation
of magnetic fields in massive stars. Discovery of strong magnetic
fields in roughly 10\% of massive stars. X-ray line emission from magnetically
confined winds. Interaction of magnetic fields and radiation-driven
winds. Effects by magnetic fields on convective structures.
Exploration of the turbulent nature of the general field of the Sun.
Magnetic cycles at varying activity levels. Polarized radiation,
gyrosynchrotron and X-ray emission. Deriving and interpreting
Zeeman-Doppler images of stellar surfaces. Hanle effect diagnostics in
stellar environments.

{\it Radiative transfer and emergent stellar spectra.} Effects on
atmospheric structure by deviations from local thermodynamic
equilibrium (non-LTE). Multidimensional radiative transfer. 
3D non-LTE spectral line formation. 
Radiative hydrodynamics. Origin and transfer of polarized light. Theory of
scattering of polarized light by atoms and molecules, particularly for
understanding the second solar spectrum. Numerical methods in
radiative transfer. Scattering mechanisms in circumstellar disks.
Impact of optically thick clumps in stellar winds on line and
continuum formation (UV and X-rays).
Atomic and molecular opacities (line and continuous). 

{\it Spectral lines and their formation.} Line formation in convective
atmospheres. Explanation of classical micro-/macroturbulent velocities as
convective and oscillatory Doppler shifts. 
Wavelength shifts and spectral line profiles as signatures of convection. 
Non-LTE radiative transfer. 
3D LTE spectral line formation. Full 3D non-LTE radiative transfer
calculations with realistic atomic models. 
Hydrogen lines as stellar thermometers. 
Fe excitation and ionization balance as temperature and gravity indicators. 
Automated and accurate abundance analyses for large-scale spectroscopic surveys. 
Detection
of strong broadening of metal lines in OB supergiants, and
interpretation in terms of supersonic `macroturbulence'. Spectra of
rapidly rotating stars viewed pole-on and equator-on. Non-LTE effects
in permitted and forbidden lines. Explanation of N\,{\sc iii}
$\lambda\lambda$ 4634-4640-4642 emission in O-stars due to wind
effects. Line formation of optical C\,{\sc iii} lines in massive stars.
NLTE IR-diagnostics of massive hot stars, particularly those
close to the Galactic Centre. X-ray line emission from line-driven
winds in massive stars. Atomic and quantum processes affecting
spectral lines. Databases for spectral lines. Atlases of synthetic
spectra.

{\it Forbidden lines and maser emission.} Molecules in atmospheres of
cool giant stars. Effects of fluorescence. Permitted and forbidden
lines from shocked atmospheres of pulsating giants. Maser and laser
emission from stellar envelopes.

{\it Chemical abundances.} 
Precise abundance measurements in BA
supergiants. N abundances for massive stars as a function of rotation,
which challenge present stellar evolution models. Abundance anomalies.
Solar chemical composition. 
3D and/or non-LTE based stellar abundance analyses. 
Chemical compositions of stars from large-scale spectroscopic surveys. 
Neutron-capture elements in solar-type and low-metallicity stars. 
Li, C and O abundances in exoplanet host stars to probe planetary properties. 
Stellar chemical signatures of planet formation. 
Lithium isotopes in metal-poor stars. 
Nucleosynthetic signatures of the first stars. 
Hydrodynamical models of metal-poor stars. Depletion of light elements
through convection and mixing processes: diffusion, rotation, turbulence etc. 
Pollution of atmospheres by interstellar
dust. r- and s-process elements. Chemical stratification in stable
atmospheres. Coronal versus photospheric abundances. Chemical
inhomogeneities and pulsation.

{\it Molecules.} Theory of the molecular Paschen-Back effect,
scattering and Hanle effect in molecular lines in the Paschen-Back
regime.
Molecular linelists based on realistic quantum mechanical calculations.
Molecule and dust formation in brown dwarfs and planetary atmospheres.
Non-equilibrium chemistry. 
Non-LTE radiative transfer for molecules.
Molecular lines as abundance indicators in the Sun and late-type stars. 
Molecules to infer stellar parameters of M dwarfs.

\subsection{Stellar structure}
\noindent
{\it Structures across stellar disks.} Doppler mapping of starspots.
Radii and oblateness at different wavelengths for giant stars.
Gravitational micro-lensing and transits to test model atmospheres. Interaction
between rotation and pulsation. Doppler tomography of stellar
envelopes.
Limb-darkening. 

{\it Stellar coronae.} Coronal heating mechanisms (quiescent and
flaring). Effects of age and chemical abundance. Multicomponent
structure. Coronae in also low-mass stars and brown dwarfs.
Diagnostics through X-ray spectroscopy and radio emission. Stellar
winds and mass loss. 
First Ionization Potential effect. 

{\it Dynamic outer atmospheres.} Multi-component
radiation and dust-driven winds. Mass loss from pulsating red giants and AGB stars. 
Effects of mass flows on the ionization structure. Coronal mass
ejections. Instabilities in hot-star winds. X-ray emission.

{\it Dust, grains, and shells.} 
Dust nucleation and condensation. 
Interplay between convection, pulsations, dust formation and mass loss in AGB stars. 
Formation of stellar dust shells.
Grains in the atmospheres of red giants, and in T Tauri stars.

\subsection{Different classes of objects}
\noindent
{\it Stellar parameters of massive stars.} Significant downscaling of
effective temperature scale of OB-stars, due to line-blanketing and
mass-loss effects. Rotation rates for massive stars, as a function of
metallicity. VLT-FLAMES survey of massive stars: 86 O-stars and 615
B-stars in 8 different clusters (Milky Way, LMC, SMC) observed and
analyzed. VLT-FLAMES Tarantula Survey: High resolution spectroscopy
and analysis of more than 1000 massive stars in the Tarantula Nebula,
incl. 300 O-type stars. First explanation of Vz stars. Stellar
parameters/abundances of Red Supergiants.

{\it Stellar parameters of late-type stars.} 
Testing different methods for stellar parameter estimations using benchmark stars
with almost model-independent parameters (e.g. interferometry, parallaxes,
binaries, asteroseismology). 
3D/non-LTE based excitation and ionization balance of Fe and other metals to
infer effective temperatures and surface gravities.
H lines as stellar thermometers: effects of 3D stellar atmospheres and non-LTE line formation. 
Line depth ratios. 
Strictly differential line-by-line analysis to achieve extremely precise relative stellar parameters and abundances. 
Molecules as temperature and gravity indicators in giants.
Wings of strong, pressure-dampened lines as gravity indicators. 
Colour-effective temperature calibration for broad-band and intermediate-band photometry.
Infrared flux method. 
Automated and accurate stellar parameter estimations for large-scale spectroscopic surveys. 
Global spectrum fitting. 
Data-driven stellar parameter estimations without stellar or spectral modelling. 
Improved stellar parameters for metal-poor stars. 
Predicted micro- and macroturbulent velocities from 3D stellar surface convection models. 
Stellar age signatures imprinted in stellar spectra (e.g. H lines, C/N ratio). 

{\it Stellar winds and mass-loss.} Inhomogeneities in stellar winds
affect derived mass-loss rates. Weak-winds from late O-/early
B-dwarfs, which display significantly lower mass-loss rates than
theoretically expected. Self-consistent models for winds from Wolf-
Rayet stars. Theoretical mass-loss rates from optically thick winds
close to the Eddington limit. Empirical mass loss-metallicity relation
for O-stars. Empirical mass-loss rates from X-ray line emission in
wind-embedded shocks. Stellar winds at very low abundances (First
Stars). Continuum driven winds for super-Eddington stars.
Wind-diagnostics by linear polarization variability. 
Dust-driven and molecular-driven winds of red giants and AGB stars. 
Interplay between convection, pulsations and stellar winds. 
Predicted mass loss rates of late-type stars. 

{\it Pulsating stars and helio-/asteroseismology.} 
Conflict between solar abundances inferred from helioseismology and 
solar spectroscopy using 3D solar atmosphere models and non-LTE line formation.
Asteroseismic scaling relations to infer stellar radii, masses, gravities and ages. 
Excitation of solar-like oscillations in stars by stellar surface convection. 
Classically variable
stars, and `ordinary' solar-type ones. Evidence for opacity-driven
gravity-mode oscillations in periodically variable B-type supergiants.
Inverting observed pressure-mode frequencies into atmospheric and interior
structure. Mass-loss mechanisms in pulsating stars. Effects of rapid
rotation on pulsation. Potential relation between oscillations
and `macro-turbulence' in massive stars.

{\it Binary stars.} Atmospheric structure and magnetic dynamos in
common-envelope binaries. Role of binarity on mass loss. Tidal
effects. Non-LTE effects by illumination from the component.
Reflection effects in close binaries. Colliding stellar winds.
Prominent role of binarity in massive stars (50-70\% of massive stars
in binaries). Test of stellar parameters of massive and low-mass stars from eclipsing
binaries.

{\it New classes of very cool stars.} Dust, clouds, weather, and
chemistry in brown dwarfs (spectral types L, T, Y). Cloud clearings and hot-spots. Magnetic
activity. The effective temperature scale. Molecular line and
continuum opacities. Transition between extrasolar giant planets and
ultracool brown dwarfs.

{\it White dwarfs and neutron stars.} Radiative transfer in magnetized
white-dwarf atmospheres. Stokes-parameter imaging of white dwarfs.
Molecular opacities in white dwarfs. 3D convection modelling of white dwarfs.
Broad-band polarization in
molecular bands in white dwarfs. Atmospheres and spectra of neutron
stars. Effects of vacuum polarization and accretion around magnetized
neutron stars.

{\it Special objects.} Central stars of planetary nebulae. Winds of
central stars as a tool to constrain their masses. Population
II and III stars of extremely low metallicity. Protostars. Accretion disks
and coronal activity in young stars.

{\it Interaction with exoplanets.} Characteristics of stars hosting
exoplanets. 
Stellar chemical signatures of planet formation.
Stellar Li, C and O abundances to infer planetary properties. 
Effects of planets on the
atmospheres of evolved red giants. 

\subsection{Development of techniques}
\noindent
{\it Computational techniques.} Parallel (super)computing to simulate
convective surface regions, and throughout complete stars. 2- and 3-D
3D radiative magneto-hydrodynamical models of stellar surface convection and atmospheres.
3D non-LTE radiative transfer and spectrum formation. 
Polarized radiative transfer. 
NLTE unified models including winds for hot stars. 
Neural networks and
machine-learning algorithms. Analysis of stellar spectra using genetic
algorithms. Automated analysis of stellar spectra using libraries of
synthetic spectra. Data-driven stellar parameter estimations without stellar or
spectral models. Preparing for the widely distributed network of
computational tools and shared databases being developed for the
forthcoming computing infrastructure GRID.

\subsection{Applications of stellar atmospheres}
\noindent
Besides their study per se, stars are being used as probes for other
astrophysical problems:

{\it Exoplanets.} Variable wavelength shifts in stellar spectra serve
as diagnostics for radial-velocity variations induced by orbiting
exoplanets. Atmospheric modeling can indicate which spectral features
are suitable as such probes, and which should be avoided due to their
sensitivity to intrinsic stellar activity.
Stellar granulation, oscillations and activity impacts on the radial velocity precision,
specially when reaching for sub-m/s level. 

{\it Chemical evolution in the Galaxy.} How accurately observations of
stellar spectral features can be transformed into actual chemical
abundances depends sensitively on the sophistication of the stellar
model atmospheres and radiative transfer modelling for predicting
the emergent stellar spectra. 

{\it Kinematics of the Galaxy.} The astrometric Gaia space mission
measures radial velocities and stellar parameters/abundances 
for huge numbers of stars. Model atmospheres
are used to identify suitable spectral features for such measurements
in different classes of stars and predict intrinsic convective blue-shifts. 

{\it Chemical evolution of Galaxies.} Accurate measurements of
metallicity gradients in various galaxies based on BA supergiants.
Late-type stars as probes of stellar, galactic and cosmic evolution. 
First stars and their immediate successors as relics from the first billion years after the Big Bang. 
Chemistry in external galaxies based on red supergiants. 

{\it Evolution of First Stars.} Effects from winds may be stronger
than expected, due to fast rotation, continuum driving or
self-enrichment. Strong mass-loss offers possibility to avoid
Pair-Instability SNe.

{\it Distance scales.} Flux-weighted gravity-luminosity relationship
as a tool to derive independent, precise extragalactic distances from
A-supergiants and Red Supergiants.

{\it Galaxies and cosmology.} Stars are the main observable component
of galaxies, and population synthesis for galaxies utilize model
atmospheres to interpret observations. Cosmological origin of the
lowest-metallicity stars. Have any Pop. III star survived from redshifts $>10$ to the present time?
Cosmological Li problems: $^7$Li and $^6$Li.

\section{Useful Web links} 
\noindent
The following collection of links (in alphabetical order) provides introductions and overviews
of several significant subfields of the physics of stellar
atmospheres.

\subsection{Calculating atmospheric models and spectra}
\noindent
ATLAS, SYNTHE, and other model grids:
$<$\href{http://kurucz.harvard.edu}{\tt kurucz.harvard.edu}$>$ \\
CCP7 - Collaborative Computational Project: $<$\href{http://ccp7.dur.ac.uk}
{\tt ccp7.dur.ac.uk}$>$\\
CLOUDY - photoionization simulations: $<$\href{http://trac.nublado.org}
{\tt trac.nublado.org}$>$\\
CMFGEN - stellar atmosphere code: \\
\hspace*{0.4cm} $<$\href{http://kookaburra.phyast.pitt.edu/hillier/web/CMFGEN.htm}
{\tt kookaburra.phyast.pitt.edu/hillier/web/CMFGEN.htm}$>$\\
CO5BOLD - 3D hydrodynamical models of stellar surface convection, atmospheres and spectra:\\
\hspace*{0.4cm}$<$\href{http://www.astro.uu.se/~bf/co5bold_main.html}
{\tt http://www.astro.uu.se/\~{}bf/co5bold\_main.html}$>$\\
INSPECT - non-LTE spectral line formation for late-type stars:\\ 
\hspace*{0.4cm}$<$\href{http://inspect-stars.com/}{\tt http://inspect-stars.com/}$>$\\
MARCS, model grids: $<$\href{http://marcs.astro.uu.se}{\tt marcs.astro.uu.se}$>$ \\
MOOG - stellar spectrum synthesis code: $<$\href{http://www.as.utexas.edu/~chris/moog.html}{\tt http://www.as.utexas.edu/\~{}chris/moog.html}$>$ \\
MULTI - non-LTE radiative transfer: $<$\href{http://folk.uio.no/matsc/mul22/}
{\tt folk.uio.no/matsc/mul22/}$>$\\
PANDORA - atmospheric models and spectra:\\
\hspace*{0.4cm}$<$\href{http://cfa.harvard.edu/~avrett/pandora.lis.copy}
{\tt cfa.harvard.edu/\~{}avrett/pandora.lis.copy}$>$\\
PHOENIX - stellar and planetary atmosphere code:\\
\hspace*{0.4cm}$<$\href{http://www.hs.uni-hamburg.de/EN/For/ThA/phoenix/}
{\tt hs.uni-hamburg.de/EN/For/ThA/phoenix/}$>$\\
PoWR: The Potsdam Wolf-Rayet Models, grid of synthetic spectra:\\
\hspace*{0.4cm}
$<$\href{http://www.astro.physik.uni-potsdam.de/~wrh/PoWR/powrgrid1.html}
{\tt astro.physik.uni-potsdam.de/\~{}wrh/PoWR/powrgrid1.html}$>$\\
RH - 1D, 2D, 3D non-LTE stellar spectrum synthesis code: \\
$<$\href{http://www4.nso.edu/staff/uitenbr/rh.html}{\tt http://www4.nso.edu/staff/uitenbr/rh.html}$>$ \\
STAGGER - 3D hydrodynamical models of stellar surface convection, atmospheres and spectra:\\
\hspace*{0.4cm}$<$\href{http://www.stagger-stars.net}
{\tt http://www.stagger-stars.net}$>$\\
STARLINK - theory and modeling resources:\\
\hspace*{0.4cm} $<$\href{http://www.astro.gla.ac.uk/users/norman/star/sc13/sc13.htx}
{\tt astro.gla.ac.uk/users/norman/star/sc13/sc13.htx}$>$\\
Synthetic spectra overview: $<$\href{http://www.am.ub.es/~carrasco/models/synthetic.html}
{\tt am.ub.es/\~{}carrasco/models/synthetic.html}$>$\\
TLUSTY/SYNSPEC - model atmospheres: $<$\href{http://nova.astro.umd.edu}
{\tt nova.astro.umd.edu}$>$\\
Tuebingen: Stellar atmosphere code, grid of models, etc.: \\
\hspace*{0.4cm} 
$<$\href{http://www.physik.uni-tuebingen.de/institute/astronomie-astrophysik/
institut/astronomie/kontakt/mitarbeiter/thomas-rauch.html} 
{\tt physik.uni-tuebingen.de/institute/astronomie-astrophysik/} \\
\hspace*{0.5cm} {\tt institut/astronomie/kontakt/mitarbeiter/thomas-rauch.html}$>$\\
WM-Basic: unified hot star model atmospheres incl. consistent wind structure:\\
\hspace*{0.4cm} $<$\href{http://www.usm.uni-muenchen.de/people/adi/adi.html}
{\tt usm.uni-muenchen.de/people/adi/adi.html}$>$\\

\subsection{Research groups or individual researchers}
\noindent
AIP Potsdam: Stellar convection:\\
\hspace*{0.4cm} $<$\href{http://www.aip.de/de/forschung/forschungsschwerpunkt-kmf/cosmic-magnetic-fields/stellar/stellar-physics/stellar-convection}
{\tt aip.de/de/forschung/forschungsschwerpunkt-kmf/} \\
\hspace*{0.5cm} {\tt cosmic-magnetic-fields/stellar/stellar-physics/stellar-convection}$>$\\
F. Allard: PHOENIX model atmospheres and radiative transfer: \\ 
\hspace*{0.4cm}$<$\href{http://perso.ens-lyon.fr/france.allard/}{\tt http://perso.ens-lyon.fr/france.allard/}$>$ \\
M. Asplund: Stellar atmospheres and spectroscopy: \\ 
\hspace*{0.4cm}$<$\href{http://www.mso.anu.edu.au/~martin}{\tt http://www.mso.anu.edu.au/\~{}martin}$>$ \\
G. Basri: Brown dwarfs: $<$\href{http://w.astro.berkeley.edu/~basri/bdwarfs/}
{\tt w.astro.berkeley.edu/\~{}basri/bdwarfs/}$>$\\
M. Bergemann: Non-LTE radiative transfer: $<$\href{http://www.mpia.de/~bergemann/}{\tt http://www.mpia.de/\~{}bergemann/}$>$\\
CO5BOLD: B. Freytag, H. Ludwig, M. Steffen et al.: 3D hydrodynamical models of stellar surface convection, atmospheres and spectra:\\
\hspace*{0.4cm}$<$\href{http://www.astro.uu.se/~bf/co5bold_main.html}
{\tt http://www.astro.uu.se/\~{}bf/co5bold\_main.html}$>$\\
A. Collier Cameron: Starspots and magnetic fields on cool stars:\\
\hspace*{0.4cm} $<$\href{http://star-www.st-and.ac.uk/~acc4/coolpages/imaging.html}
{\tt star-www.st-and.ac.uk/\~{}acc4/coolpages/imaging.html}$>$\\
P. Crowther: Hot Lumionous Star Research Group:\\
\hspace*{0.4cm} $<$\href{http://www.pacrowther.staff.shef.ac.uk/science.html}
{\tt pacrowther.staff.shef.ac.uk/science.html}$>$\\
J. F. Donati: Magnetic fields of non degenerate stars:\\
\hspace*{0.4cm} $<$\href{http://www.ast.obs-mip.fr/users/donati/}
{\tt ast.obs-mip.fr/users/donati/}$>$\\
D. Dravins: Stellar surface structure and more:
$<$\href{http://www.astro.lu.se/~dainis/}{\tt astro.lu.se/\~{}dainis/}$>$\\
D. F. Gray: Stellar rotation, magnetic cycles, velocity fields:
$<$\href{http://www.astro.uwo.ca/~dfgray/}{\tt astro.uwo.ca/\~{}dfgray/}$>$\\
P. Hauschildt: PHOENIX model atmospheres and radiative transfer: \\ 
\hspace*{0.4cm}$<$\href{http://hobbes.hs.uni-hamburg.de/}{\tt http://hobbes.hs.uni-hamburg.de/}$>$ \\
C. Helling: atmospheres of brown dwarfs and (exo-)planets: \\ 
\hspace*{0.4cm}$<$\href{https://leap2010.wp.st-andrews.ac.uk/}{\tt https://leap2010.wp.st-andrews.ac.uk/}$>$ \\
S. H\"ofner: dynamical atmospheres and winds of AGB stars: \\ 
\hspace*{0.4cm}$<$\href{http://www.astro.uu.se/~hoefner/}{\tt http://www.astro.uu.se/\~{}hoefner/}$>$ \\
M. Jardine: Stellar coronal structure:\\
\hspace*{0.4cm} $<$\href{http://www-star.st-and.ac.uk/~mmj/Research\_cool\_stars.html}
{\tt www-star.st-and.ac.uk/\~{}mmj/Research\_cool\_stars.html}$>$\\
S. Jeffery: Stellar model grids, hot stars: 
$<$\href{http://star.arm.ac.uk/~csj/}{\tt star.arm.ac.uk/\~{}csj/}$>$\\
F. Kupka: 3D stellar convection modelling, atomic data: \\
\hspace*{0.4cm}$<$\href{http://www.mat.univie.ac.at/~kupka/}{\tt http://www.mat.univie.ac.at/\~{}kupka/}$>$\\
R. Kudritzki: Hot Stars and Winds, Extragalactic Stellar Astronomy:\\
\hspace*{0.4cm} $<$\href{http://www.ifa.hawaii.edu/~kud/kud.html}{\tt ifa.hawaii.edu/\~{}kud/kud.html}$>$\\
K. Lind: Non-LTE radiative transfer: 
$<$\href{www.astro.uu.se/~klind/index/Welcome.html}{\tt www.astro.uu.se/\~{}klind/index/Welcome.html}$>$\\
J.L.Linsky: Cool stars, stellar chromospheres and coronae:\\
\hspace*{0.4cm} $<$\href{http://jilawww.colorado.edu/~jlinsky/}{\tt jilawww.colorado.edu/\~{}jlinsky/}$>$\\
D. Montes et al.: Libraries of stellar spectra:\\
\hspace*{0.4cm} $<$\href{http://pendientedemigracion.ucm.es/info/Astrof/invest/actividad/spectra.html}
{\tt pendientedemigracion.ucm.es/info/Astrof/invest/actividad/spectra.html}$>$\\
Munich Hot star group (A. Pauldrach and J.Puls), lecture notes and more:\\
\hspace*{0.4cm}$<$\href{http://www.usm.uni-muenchen.de/people/adi/wind.html}
{\tt usm.uni-muenchen.de/people/adi/wind.html}$>$,\\
S. Owocki: Theory of line-driven winds, hydrodynamics, rotation,
magnetic fields:\\
\hspace*{0.4cm} $<$\href{http://www.bartol.udel.edu/~owocki/}{\tt bartol.udel.edu/\~{}owocki/}$>$\\
N. Przybilla: NLTE atmospheres of massive stars, extragalactic stellar astronomy:\\
\hspace*{0.4cm} $<$\href{http://www.sternwarte.uni-erlangen.de/~przybilla/research.html}
{\tt sternwarte.uni-erlangen.de/\~{}przybilla/research.html}$>$  (Note: site to be moved)\\
R. J. Rutten: Lecture notes: Radiative transfer in stellar atmospheres and more:\\
\hspace*{0.4cm} $<$\href{http://www.staff.science.uu.nl/~rutte101/Course\_notes.html}
{\tt staff.science.uu.nl/\~{}rutte101/Course\_notes.html}$>$\\
C. Sneden - MOOG stellar spectrum synthesis code: 
$<$\href{http://www.as.utexas.edu/~chris}{\tt http://www.as.utexas.edu/\~{}chris}$>$ \\
STAGGER: M. Asplund, R. Collet, Z. Magic, \AA . Nordlund et al.: 3D hydrodynamical models of stellar surface convection, atmospheres and spectra:
$<$\href{http://www.stagger-stars.net}{\tt http://www.stagger-stars.net}$>$\\
P. Stee: Be-star atmospheres and circumstellar envelopes:\\
\hspace*{0.4cm} $<$\href{https://www-n.oca.eu/stee/page1/page1.html}
{\tt https://www-n.oca.eu/stee/page1/page1.html}$>$\\
R. F. Stein: Convection simulations \& radiation hydrodynamics:\\
\hspace*{0.4cm} $<$\href{http://www.pa.msu.edu/~steinr/research.html\#convection}
{\tt pa.msu.edu/\~{}steinr/research.html\#convection}$>$\\
R. Townsend: Astrophysics of massive stars:\\
\hspace*{0.4cm} $<$\href{http://www.astro.wisc.edu/~townsend/static.php?ref=research}
{\tt astro.wisc.edu/\~{}townsend/static.php?ref=research}$>$\\
H. Uitenbroek: Non-LTE radiative transfer: $<$\href{http://www4.nso.edu/staff/uitenbr/}{\tt http://www4.nso.edu/staff/uitenbr/}$>$\\
Vienna: Stellar atmospheres and pulsating stars: 
$<$\href{http://www.univie.ac.at/asap}{\tt univie.ac.at/asap}$>$\\

{\hfill Joachim Puls and Martin Asplund}

{\hfill {\it President and Past President of the Commission}}

{}

\end{document}